\documentstyle[12pt]{article}

\advance\textheight by \topskip

\begin{document}
%

\hfill
\renewcommand{\thefootnote}{\fnsymbol{footnote}}
NIKHEF-H/94-30\footnote{The complete postscript file of this preprint
is available via anonymous ftp at nikhefh.nikhef.nl as
/pub/preprints/H94-30.ps.gz }

\addtocounter{footnote}{-1}
\renewcommand{\thefootnote}{\arabic{footnote}}
\hfill
September 1994

\begin{center}
{\huge \bf The  large quark mass expansion of 
$\Gamma (Z^{0}\rightarrow hadrons)$ and 
$\Gamma(\tau^{-} \rightarrow \nu_{\tau}+hadrons)$ in the
order $\alpha_{s}^3$
 }\\ [8mm]
S.A. Larin\footnote{On leave from the Institute for Nuclear
Research (INR) of the Russian Academy of Sciences, Moscow 117312.},
T. van Ritbergen, J.A.M. Vermaseren \\ [3mm]
NIKHEF-H, P.O. Box 41882, \\ 1009 DB, Amsterdam \\
\end{center}
\begin{abstract}
We present the analytical $\alpha_{s}^3$ correction to the  $Z^{0}$ decay rate 
into hadrons. We calculate this correction up to (and including) 
terms of the order $(m_Z^2/m_{top}^2)^3$ in the large top quark mass expansion.
We rely on the technique of the large mass expansion of individual Feynman
diagrams and treat its application in detail.
We convert the obtained results of six flavour QCD to the results in
the effective theory with five active flavours, checking the
decoupling relation of the QCD coupling constant. We also 
derive the large charm quark mass expansion of the semihadronic $\tau$ 
lepton decay rate in the $\alpha_{s}^3$ approximation.

\end{abstract}
\newpage
\section{Introduction}
Precision measurements of the $Z^{0}$ decay rate into hadrons at LEP
\cite{lep1} provide precise means to extract the QCD coupling constant
from experiment. 
This is a very clean process from a theoretical point of view since its
calculation can be reduced to the calculation of the Z boson propagator
within the standard model.
The status of electroweak
corrections to $Z^{0}$ decay can be found in ref. \cite{zfitter}.
Now the calculational techniques of Feynman diagrams have advanced 
so far that the calculation of the $\alpha_{s}^3$ order
(=4 loop approximation of the Z boson propagator) is feasible.
The $\alpha_{s}^3$ approximation to the $Z^{0}$ decay rate into hadrons
is important for an accurate determination of the QCD coupling constant
$\alpha_{s}$, or
equivalently the fundamental scale of QCD, $\Lambda_{QCD}$.

The hadronic $Z^{0}$ decay rate is a sum of vector and 
axial vector contributions of which
the vector contribution is known
to order $\alpha_{s}^3$ from the calculation of $\sigma_{tot}(e^{+}e^{-}
\rightarrow \gamma \rightarrow hadrons)$ \cite{gorishny2}.
This calculation was performed in the approximation of effective QCD
with five massless quarks which involved the calculation of only
massless diagrams.
The correctness of this calculation is strongly supported by
\cite{broadhurst2} where the non-trivial connection between the result
\cite{gorishny2} and the $\alpha_{s}^3$ approximation \cite{gls}
 to deep inelastic sum rules was established.

The calculation of the axial vector part of the hadronic $Z^{0}$ decay rate
is more involved than that of the vector part.
This is because the heavy quark does not decouple in the axial vector part
and one cannot avoid to calculate massive diagrams, even in the leading
order of the large mass expansion.
The axial vector part was calculated
to order $\alpha_{s}^2$ in \cite{kniehl1} and confirmed in \cite{chetyrkin1}
where the operator product expansion technique was used to sum up the massive
logarithmic terms.
The $Z^{0}$ decay into 3 gluons in order  $\alpha_{s}^3$
has been calculated in \cite{bij1}. The $\alpha_{s}^3$ 
correction to the axial vector part of the hadronic $Z^{0}$ decay rate
 in the leading order of the large top mass
expansion was presented in \cite{we,they}. 
In this paper we elaborate the details of the calculation \cite{we} and
extend the large mass expansion of both the vector and axial vector
parts to the order $(m_Z^2/m_{top}^2)^3$. This calculation allows us
also to check the decoupling mechanism at the next-next-to-leading order.

Another prominent process (beside Z boson decay) 
for the extraction of $\alpha_{s}$ from
experiment is semihadronic $\tau$ lepton decay. 
In the last section of our article we convert the obtained result
for $\Gamma (Z^{0}\rightarrow hadrons)$ to the large charm quark mass
expansion for $\Gamma(\tau^{-} \rightarrow \nu_{\tau}+hadrons)$ 
in the order $\alpha_{s}^3$.

\section{Preliminaries}
For the $Z^{0}$ decay rate into hadrons,
the quantity to be determined is the squared matrix element summed over
all final hadronic states.
One can express this quantity as the imaginary part of a
current correlator in the standard way
\begin{equation} \sum_{h} <0| J^{\mu} |h> <h| J^{\nu}|0> \hspace{.15cm} =
\hspace{.05cm} 2 Im \Pi^{\mu \nu} ,\end{equation}
\begin{equation} \label{correlator} 
\Pi^{\mu \nu} = i \int d^{4}z \mbox{\bf e}^{i q\cdot z}
<0| T( J^{\nu}(z) J^{\mu}(0) ) |0> \hspace{.15cm}=
 -g^{\mu \nu}q^2 \Pi_{1}(q^{2})
-q^{\mu}q^{\nu}\Pi_{2}(q^{2}). \end{equation}
Here $ J^{\mu}=\frac{g}{2 \cos_{\theta_W}}\sum_{i=1}^{6}
\overline{\psi_{i}}\gamma^{\mu} \left( g_{V}^{i}-g_{A}^{i}\gamma^{5} \right)
 \psi_{i} $
is the neutral weak quark current coupled to the $Z^{0}$ boson in the 
Lagrangian of the Standard Model, where we use the notations as  
 given in \cite{properties}
$g_{V}^i \equiv t_{3L}(i)-2q_i \sin^2 \theta_{W}$
and  $g_{A}^i \equiv t_{3L}(i)$. 

The hadronic $Z^{0}$ decay width is expressed as
\begin{equation} \label{impi}
 \Gamma_{had} \equiv \Gamma_{had}^{V} + \Gamma_{had}^{A} 
= m_{Z}Im\Pi_{1}(m^{2}_{Z}+ i\hspace{.03cm} \epsilon)
\end{equation}
with the indicated decomposition into vector and axial vector parts imposed
by the structure of the neutral current.
We will calculate $Im \Pi_{1}$ in the order $g^2 \alpha_s^3$.
It is a calculation within perturbative QCD except for two weak current vertex
insertions (i.e. the weak current is considered as an external current for
QCD).

Throughout this paper we use dimensional regularization \cite{dimreg} 
in $D = 4-2\varepsilon$ space-time dimensions
and the standard modification of the minimal subtraction scheme
\cite{ms}, the $\overline{MS}$ scheme \cite{msbar}.
For the treatment of the $\gamma_{5}$
matrix in dimensional regularization we use the technique  described
in \cite{larin2} which is based on the original definition of $\gamma_{5}$
in \cite{dimreg}. We work in the approximation of 5 massless quark flavours
and the top quark mass large compared to the $Z^{0}$ mass. 
We should stress that the top quark does not
decouple \cite{nodecouple} from the axial vector part due to diagrams 
of the axial anomaly type. 

It is convenient to split the vector and axial vector contribution in 
non-singlet and singlet parts
\begin{equation}
\Gamma_{had}^{V} = \Gamma_{had}^{V,NS} + \Gamma_{had}^{V,S},
\hspace{1cm}
\Gamma_{had}^{A} = \Gamma_{had}^{A,NS} + \Gamma_{had}^{A,S}.
\end{equation}
The non-singlet parts come from Feynman diagrams where both weak current
 vertices are located in one fermion loop.The singlet contributions come
from diagrams where each weak current vertex is located in a separate
fermion loop. The massive non-singlet diagrams are presented in
figure 1, the singlet diagrams are presented in figures 2,3.

The $\alpha_{s}^3$ approximation for the
vector part in effective QCD with 5 active massless quark flavours
in the $\overline{MS}$ scheme was calculated in \cite{gorishny2}
(in the leading order of the large top quark mass expansion). 
This calculation used the fact that the top quark decouples for the
vector part in the leading order of the large quark mass expansion.
Therefore within effective 5 flavour QCD this calculation 
involved only massless diagrams and the result reads
\[
\Gamma_{had}^{V,NS} = \frac{G_{F}m_{Z}^{3}}{2\pi\sqrt{2}} \sum_{i=1}^{5}
(g_{V}^i)^{2} \left[ 1+\frac{\alpha_{s}^{(5)}}{\pi}
+1.40923 \left( \frac{\alpha_{s}^{(5)}}{\pi} \right)^{2}
-12.76706 \left( \frac{\alpha_{s}^{(5)}}{\pi} \right)^{3} \right] \]\nopagebreak 
\begin{equation} \label{gammavec}
\Gamma_{had}^{V,S} = 
  \frac{G_{F}m_{Z}^{3}}{2\pi\sqrt{2}}
  \left( \sum_{i=1}^{5}g_{V}^i \right)^{2}
  \left[ -0.41318 \left( \frac{\alpha_{s}^{(5)}}{\pi} \right)^{3} \right] 
\end{equation}
with the Fermi constant $G_{F} = \frac{g^{2}\sqrt{2}}{8 \cos^2(\theta_W)
 m_{Z}^{2} }$.
Here $\alpha_{s}^{(5)}(m_{Z})$ is the coupling constant in
effective QCD with 5 active flavours.
The coupling constant of effective 5 flavour QCD, $\alpha_{s}^{(5)}$
 and the coupling constant of
full 6 flavour QCD, $\alpha_{s}^{(6)}$
 both obey the renormalization group equation
( with $n_f = 5$ for $\alpha_{s}^{(5)}$ and $n_f = 6$ for $\alpha_{s}^{(6)}$ )
\begin{eqnarray}
\label{rengroup}
\frac{\partial\, \alpha_s / \pi}{\partial \, ln\, Q^2} & = &
\beta(\frac{\alpha_s}{\pi}) \nonumber \\
& = & -\beta_0(\frac{\alpha_s}{\pi})^2 - \beta_1(\frac{\alpha_s}{\pi})^3
-\beta_2 (\frac{\alpha_s}{\pi})^4 + O(\alpha_s)^5
\end{eqnarray}
where 
\begin{eqnarray}
\beta_{0} & = & \frac{1}{4}( \frac{11}{3} C_A - \frac{4}{3} T_F n_f )
 \nonumber \\
 \beta_{1} & = & \frac{1}{16}
( \frac{34}{3}C_A^2 - 4 C_F T_F n_f -\frac{20}{3} C_A T_F n_f
 )  \nonumber \\
 \beta_{2} & = &\frac{1}{64} ( \frac{2857}{54} C_A^3
 +2 C_F^2 T_F n_f - \frac{205}{9} C_F C_A T_F n_f - \nonumber \\
 & & - \frac{1415}{27} C_A^2 T_F n_f 
 + \frac{44}{9} C_F T_F^2 n_f^2 
  + \frac{158}{27} C_A T_F^2 n_f^2 )
\end{eqnarray}
The three loop QCD beta function in the $\overline{MS}$-scheme
was calculated in \cite{tvz:3beta}.
 $C_{F} = \frac{4}{3}$ and $C_{A}= 3$ are the Casimir operators 
of the fundamental and 
adjoint representation of the colour group $SU(3)$,
$T_{F} = \frac{1}{2}$ is the
trace normalization of the fundamental representation.

The solution of eq.(\ref{rengroup}) 
in the next-next-to-leading order has the standard form
\begin{eqnarray}
\label{alphaeff}
        \frac{\alpha_s}{\pi} & = &
        \frac{1}{\beta_0 \ln(\frac{Q^2}{\Lambda_{\overline{MS}}^2})}
        - \frac{\beta_1}{\beta_0^3}
        \frac{\ln\  \ln(\frac{Q^2}{\Lambda_{\overline{MS}}^2})}{
        \ln^2(\frac{Q^2}{\Lambda_{\overline{MS}}^2})} \nonumber \\
        && + \frac{1}{\beta_0^5 \ln^3(\frac{Q^2}{\Lambda_{\overline{MS}}^2})}
        \left( \beta_1^2 \ln^2\ln(\frac{Q^2}{\Lambda_{\overline{MS}}^2})
        - \beta_1^2 \ln \ \ln(\frac{Q^2}{\Lambda_{\overline{MS}}^2})
        + \beta_2\beta_0 - \beta_1^2 \right)
\end{eqnarray}
where it is understood that the scale $\Lambda_{\overline{MS}}$ 
also depends on the number of active flavours.\pagebreak

The (top quark mass dependent) relation between $\alpha_{s}^{(6)}$ and
 $\alpha_{s}^{(5)}$ is called the decoupling relation and will be discussed
later in more detail but for completeness we give the NNL order expression
\[ 
 \frac{\alpha_{s}^{(6)}(\mu)}{\pi} = \frac{\alpha_{s}^{(5)}(\mu)}{\pi}
 +\left( \frac{\alpha_{s}^{(5)}(\mu)}{\pi} \right)^{2} \frac{T_{F}}{3}
\ln(\frac{\mu^2}{m_{t}^{2}(\mu)})  \]
\begin{equation} \label{decouplingrelation}
+\left( \frac{\alpha_{s}^{(5)}(\mu)}{\pi} \right)^{3}
 \left( \frac{T_{F}^2}{9} \ln^2(\frac{\mu^2}{m_{t}^{2}(\mu)})
  +\frac{5C_A T_F-3 C_F T_F}{12} \ln(\frac{\mu^2}{m_{t}^{2}(\mu)})
 +\frac{13}{48} T_F C_F  -\frac{2}{9} T_F C_A \right)
+ O(\alpha_{s}^{4}) .
\end{equation}
where $\mu$ is the renormalization scale and $m_{t}(\mu)$ is the
top quark mass in the $\overline{MS}$ scheme. Please note that the
 term $\frac{13}{48} T_F C_F$ that we found is slightly
different from the one in ref. \cite{bernreuther}.
Substituting expression (\ref{alphaeff}) for $\alpha^{(5)}_s$ and 
$\alpha^{(6)}_s$ one can find the connection between
 $\Lambda^{(6)}_{\overline{MS}}$ and $\Lambda^{(5)}_{\overline{MS}}$
via $m_{t}(\mu)$.

$m_{t}(\mu)$ obeys the renormalization group equation
\begin{eqnarray}
\frac{\partial\, \ln(m_{t}(\mu)) }{\partial \, \ln (\mu^2)} & = &
- \gamma_m(\alpha_{s}) \nonumber \\ 
& = & - \gamma_{0} (\frac{\alpha_s}{\pi})
 -  \gamma_1 (\frac{\alpha_s}{\pi})^2 
 -  \gamma_2 (\frac{\alpha_s}{\pi})^3
 + O(\alpha_s)^4
\end{eqnarray}
where 
\begin{eqnarray}
\gamma_{0} & = & \frac{1}{4}3 C_F \nonumber \\
 \gamma_{1} & = & \frac{1}{16}
( \frac{3}{2}C_F^2 + \frac{97}{6} C_F C_A -\frac{10}{3} C_F T_F n_f  
 )  \nonumber \\
 \gamma_{2} & = &\frac{1}{64} ( \frac{129}{2} C_F^3 
 -\frac{129}{4}C_F^2 C_A + \frac{11413}{108} C_F C_A^2 + \nonumber \\ 
 & & + C_F^2 T_F n_f ( 
48 \zeta_3 - 46) + C_F C_A T_F n_f ( -48 \zeta_3
 - \frac{556}{27} ) - \frac{140}{27} C_F T_F^2 n_f^2 ) 
\end{eqnarray}
and $\zeta$ is the Riemann zeta-function.
The three loop quark mass anomalous dimension was calculated
in ref. \cite{massano}

Let us  quote the existing results for the axial vector contribution.
The axial vector
non-singlet part can be reduced to the vector case by using the 
effective anticommutation property of the $\gamma_{5}$ matrix in the
prescription that we use.
More strictly, it can be done only in the limit of massless light quarks.
Thus the non-singlet axial part coincides with the vector non-singlet part
up to a change of the weak coupling constants and reads (in the leading 
order of the large top quark mass expansion)
\begin{equation}
\Gamma_{had}^{A,NS} =
\frac{G_{F}m_{Z}^{3}}{2\pi\sqrt{2}} \sum_{i=1}^{5}
(g_{A}^i)^{2} \left[ 1+\frac{\alpha_{s}^{(5)}}{\pi}
+1.40923 \left( \frac{\alpha_{s}^{(5)}}{\pi} \right)^{2}
-12.76706 \left( \frac{\alpha_{s}^{(5)}}{\pi} \right)^{3} \right]
\end{equation} \pagebreak

Let us turn to the singlet axial vector part.
In the Standard Model quarks in a weak doublet couple with opposite
sign to the $Z^0$ boson in the axial vector part of the neutral current.
That is why the contributions from light doublets add up to zero in the massless
limit for  axial vector singlet diagrams. 
The only non-zero contribution comes from the top-bottom doublet due to
the large mass difference between top and bottom quarks.

 The axial vector singlet part in the leading
order of the large top quark mass expansion has recently been calculated
in refs. \cite{we,they}. 
 The result in the effective theory with 5 active massless quark flavours is
\[ \Gamma_{had}^{A,S} =
\frac{G_{F}m_{Z}^{3}}{2\pi\sqrt{2}} \left( g_A^{bot}\right)^2
\left[ \left( \frac{\alpha_{s}^{(5)}}{\pi} \right)^{2} \left(
 - \frac{17}{6} \right) \right. + \hspace{6cm} \]
\[\hspace{1.5cm} + \left. \left( \frac{\alpha_{s}^{(5)}}{\pi} \right)^{3} \left(
- \frac{4673}{144} + \frac{67}{12} \zeta_3 + \frac{23}{36}\pi^2
   - \frac{1}{36} \ln(\frac{m_{Z}^2}{m_{t}^2})
   - \frac{1}{6} \ln^2(\frac{m_{Z}^2}{m_{t}^2})
\right) \right] \]
\[ \hspace{.5cm} + 
\frac{G_{F}m_{Z}^{3}}{2\pi\sqrt{2}} \left( g_A^{bot} g_A^{top}\right)
\left[ \left( \frac{\alpha_{s}^{(5)}}{\pi} \right)^{2} \left(
 \frac{1}{4}
  - \ln(\frac{m_{Z}^2}{m_{t}^2}) \right) \right. + \hspace{3.5cm} \]
\[+ \left. \left( \frac{\alpha_{s}^{(5)}}{\pi} \right)^{3} \left(
  - \frac{2717}{432} + \frac{55}{12}\zeta_3
 - \frac{7}{4} \ln(\frac{m_{Z}^2}{m_{t}^2})
 - \frac{25}{12} \ln^2(\frac{m_{Z}^2}{m_{t}^2}) 
  \right) \right] \]
\[
  =
  \frac{G_{F}m_{Z}^{3}}{2\pi\sqrt{2}}
  \left( \frac{1}{4} \right)
\left[ \left( \frac{\alpha_{s}^{(5)}}{\pi} \right)^{2} \left(
- \frac{37}{12} + \ln (\frac{m_{Z}^2}{m_{t}^{2}})
\right) \right. + \hspace{3.5cm} \]
\begin{equation} \label{sin1}
+ \left. \left( \frac{\alpha_{s}^{(5)}}{\pi} \right)^{3} \left( -18.65440
+ \frac{31}{18} \ln(\frac{m_{Z}^2}{m_{t}^2})
+ \frac{23}{12}  \ln^2 (\frac{m_{Z}^2}{m_{t}^2})
\right) \right] ,
\end{equation}
where we separated the two weak coupling structures (which was not present
in refs. \cite{we,they}) and used the notation
$g_A^{bot} \equiv g_A^5$ and $g_A^{top} \equiv g_A^6$.
Here and below $m_t\equiv m_t(m_Z)$ is the $\overline{MS}$ top mass at the
scale $m_Z$.
One may relate it to the pole mass through the expression
\[ m_{t}(m_Z)=m_{pole}\left[1-\frac{\alpha_s(m_Z)}{\pi} \left(
\ln(\frac{m_Z^2}{m_{pole}^2})+\frac{4}{3}\right)+O(\alpha_s^2)\right] \]
which is known in the NNL approximation \cite{polemass}
or relate it to $m_t(m_t)$ through the expression
\[ m_{t}(m_Z)=m_t(m_t)\left[ 1-\frac{\alpha_s(m_Z)}{\pi}
\ln(\frac{m_Z^2}{m_t^2(m_t)})+O(\alpha_s^2)\right]. \]
This would correspondingly modify the coefficients of the $\alpha_{s}^{3}$
term in (\ref{sin1}) but we prefer to use 
the $\overline{MS}$ top quark mass, $m_t(\mu)$
(at $\mu = m_Z$) which is the original mass from the QCD Lagrangian.

\pagebreak
In the present paper we present the power 
suppressed top quark mass corrections
for both the vector and axial vector contributions.
The Feynman diagrams that we have to calculate to obtain the power
suppressed top quark mass corrections are given in figures \nolinebreak 1,2,3.
\[ \]
\[  \vspace{13cm} \]
\[ \parbox{14cm}{\small 
{\bf Figure 1.}
 Massive diagrams (= with top quark loops)
contributing to the vector non-singlet part,
$\Gamma_{had}^{V,NS}$ and axial
non-singlet part, $\Gamma_{had}^{A,NS}$. 
The symbol $\otimes$ is used to indicate 
an external vertex of the neutral weak vector current
 for $\Gamma_{had}^{V,NS}$ and an axial vector
vertex  for $\Gamma_{had}^{A,NS}$. 
It is understood that for each diagram at least
one fermion loop has to be a massive top quark loop 
and a loop that contains the external vertices is always a massless quark loop.
The massless diagrams that were already calculated in ref. \cite{gorishny2}
are not considered here.} \] 
\pagebreak
\[ \vspace{2cm} \]
\[ \parbox{14cm}{\small
{\bf Figure 2.} Massive diagrams (= with top quark loops)
 contributing to the vector singlet part, 
$\Gamma_{had}^{V,S}$.
The symbol $\otimes$ is used to indicate a vector current vertex.
It is understood that for each diagram at least
one fermion loop has to be a massive top quark loop.
The massless diagrams (of the same topologies)
were already calculated in ref. \cite{gorishny2}
and are not considered here. } \]

\[ \vspace{5cm} \]
\[ \parbox{14cm}{ \small
{\bf Figure 3.} Massive and massless diagrams contributing to the axial vector
singlet part, $\Gamma_{had}^{A,S}$. The symbol $\otimes$ is used to indicate 
an axial vector current vertex. 
For each fermion loop both massless quarks and massive top quarks 
are considered. } \]

\pagebreak
\section{The calculation of the massless diagrams}
In this section we will treat the calculation of the four-loop massless
diagrams from figure 3 that contribute to $\Gamma_{had}^{A,S}$.
We will illustrate the techniques by considering the most difficult
diagram (the first one on the second line in figure 3).

Since we are only interested in the structure function $\Pi_{1}(q^2)$
(see equations (\ref{correlator},\ref{impi})), we contract the diagrams with the
projector $(g^{\mu\nu}-q^{\mu}q^{\nu}/q^2)$ which reduces the diagram to
a scalar integral.
In figure 4 we present diagrammatically the renormalization of ultraviolet
divergences of this diagram.

\[  \vspace{1.4cm} \]
\[ \parbox{14cm}{\small
{\bf Figure 4.} The ultraviolet renormalization of the four-loop diagram.
The ultraviolet counterterms are presented between round brackets.} \]

At present, we do not have a technique which would allow a direct
calculation of this 4-loop diagram. But we can use the fact that we
need to know only poles in $\varepsilon$ for this diagram, since only
these pole terms generate terms containing $\ln(Q^2/\mu^2)$ which produce
non-zero imaginary parts
(logarithms come from
  $\frac{1}{\varepsilon}(\frac{\mu^{2}}{Q^2})^{\varepsilon} = 
\frac{1}{\varepsilon} + \ln(\frac{\mu^{2}}{Q^2}) + O(\varepsilon)$  and give
imaginary parts through
$\ln(-s-i\epsilon) = \ln(s) - i\pi $, where $s = -Q^2 = q^2 = m_Z^2$).
One can see from figure 4 that it is
sufficient to calculate all renormalization terms ($2^{nd}$- $4^{th}$
terms in figure 4), including the 4-loop
counterterm, in order to restore the pole terms for the diagram itself
(because the sum is finite).
All renormalization terms except the 4-loop counterterms can be directly
calculated with the help of the package MINCER \cite{mincer} written for the
symbolic manipulation program FORM \cite{form}. This package calculates
analytically 3-loop massless propagator diagrams
using the integration by part algorithm  of ref. \cite{ct}
for dimensionally regularized diagrams.

In this way we have reduced the problem of calculating the imaginary part
of the 4-loop diagram of figure 4 to the problem of calculating its
4-loop ultraviolet counterterm. The last problem can be reduced to the
calculation of 3-loop propagator type diagrams (which are
calculable with the package MINCER) using the 
infrared rearrangement method \cite{vladimirov,ckt}. This method relies
heavily on the fact that in the $\overline{MS}$-scheme ultraviolet
counterterms are polynomials in momenta and masses \cite{collinsp},
i.e. do not contain logarithms or inverse powers 
of momenta and masses. In our case
the 4-loop counterterm has dimension two which means that it is
simply proportional to $Q^2$. If we take the d'Alembertian in Q of
this counterterm we get a dimensionless quantity which is just a Laurent
series in $\varepsilon$. To obtain this Laurent series we need to calculate
counterterms of dimensionless diagrams. These  diagrams
 are produced after applying
the d'Alembertian in Q to the 4-loop diagram of figure 4. In fact, 
the application
of the d'Alembertian produces several dimensionless 4-loop diagrams after
differentiation of the lines of the original diagram; some of these diagrams
are presented in figure 5.

\[  \vspace{2cm} \]
\[ \parbox{14cm}{\small {\bf Figure 5.}
Diagrammatic representation of applying the d'Alembertian in Q
to the 4-loop counterterm. A prime on a line denotes the differentiation
of this line in its momentum. The round brackets denote 
the ultraviolet counterterms of the diagrams inside the brackets.} \]

Since counterterms of dimensionless diagrams do not depend on the external
momentum Q, we can change the route of this momentum through these
diagrams as we wish, e.g. as it is chosen in the r.h.s. of figure 5.
The choice of a new route for the external momentum can generate infrared
poles (even though the original diagram did not have
infrared divergences) which then essentially complicates the extraction
of the ultraviolet poles. For example, nullifying the momentum Q 
will nullify the whole diagram because of new infrared divergences.
The chosen momentum route in the r.h.s. of figure 5
 reduces the calculation of the 4-loop diagrams
in figure 5 to the calculation of simpler 4-loop topologies.
These topologies have the form of a
3-loop propagator type subdiagram (which can be done with the package MINCER)
inserted in a trivial one-loop topology.

The essential complication now is that the second diagram in the r.h.s.
of figure 5 contains infrared divergences.
For this diagram it is impossible to
choose the route of the external momentum in a way
that avoids infrared divergences and simultaneously reduces the calculation
of the corresponding diagram to a 3-loop propagator insertion.
In this case we need to apply the technique of the $R^{*}$
operation \cite{rstar} which allows to calculate ultraviolet counterterms
of dimensionally regularized diagrams even in the presence of infrared
singularities. It is interesting to note that the diagram of figure 4
was the only diagram that needed the application of the $R^{*}$ operation
in the calculation of $\Gamma_{had}^{A,S}$.

\pagebreak
\section{The expansion of massive diagrams}
We will now treat the large mass expansion of the individual 
massive diagrams. We calculate all integrals in Euclidean momentum space.
The general
theory of Euclidean asymptotic expansions was developed in
\cite{tkachov1,chetyrkin2}.
For practical purpose we use the techniques developed in
 \cite{gorishny1,gorishny3}.

Let's go through some simple ideas (see e.g. \cite{collins})
that generalise to a recipe of expanding individual 
dimensionally regularized diagrams.
This recipe can then also be used to expand $\overline{MS}$ renormalized
diagrams since it can be applied to each term of
the renormalized expression for a given diagram
(i.e. the expression after application of the ultraviolet R-operation
to this diagram).

A simple scalar diagram containing both massless lines and massive lines is 
shown in the l.h.s. of figure 6a.

\[ \vspace{1.6cm} \]
\[ \parbox{14cm}{ \small
{\bf Figure 6a.}
 Thick lines denote massive scalar propagators, thin lines denote
  massless scalar propagators. The symbol $\times$ indicates the insertion
of the small momentum expansion of the massive triangle. } \]
We are going to expand this diagram in a large mass which is equivalent
to the expansion in a small (in comparison to the mass) external momentum, Q.
Please note that we can not simply expand the integrand 
of the corresponding Feynman 
integral as a Taylor series in Q, because putting Q $= 0$ generates 
infrared divergences when one integrates over momentum k.
Let us therefore first consider the expansion of the massive one-loop subgraph
as a Taylor series in its external momenta, k and Q.
This expansion can be obtained by a simple Taylor expansion 
of its integrand in k and Q because it does not
generate infrared divergences when one integrates over l. 
More generally, the expansion of diagrams with only massive
propagators in terms of (small) external momenta can be safely done 
by making Taylor expansions in these momenta in the integrands.
The Taylor expansions of integrands are generated by simple expansions
of propagators in  small momenta Q
\[ \frac{1}{(P+Q)^2+M^2} =
 \frac{1}{P^2+M^2} \sum_{i=0}^{N} 
 \frac{(-2P\cdot Q - Q^2)^{i}}{(P^2+M^2)^{i}}
 + O\left(\frac{1}{(P^2+M^2)^{(N+2)}}\right) \]
where N is the desired order in the expansion.
We then add and subtract
the expansion of the massive one-loop subgraph as is indicated in
the r.h.s. of figure 6a.  

The first term in the r.h.s. of figure 6a 
 is a massless one-loop integral with the 
expanded massive subintegral inserted in the integrand and it can be
evaluated without difficulty, keeping Q finite.

The second term in the r.h.s. of figure 6a 
(the combination in the square brackets)
has a vanishing contribution from the integration region
where k is small because the behaviour of the massive subintegral 
for small k is subtracted off until the necessary order (that is determined
by the depth of the expansion).
One may then Taylor expand the 
integrand of the second term around Q $=0$ to obtain
 an integral without an external momentum
(a tadpole integral) that can also be evaluated without difficulty.
It should be noted that massless tadpole diagrams vanish in dimensional
regularization which means that after the Taylor expansions in the
integrand of the second term in Q, only the two loop massive tadpole
contribution survives. 

Finally we get the large mass expansion as it is presented in figure 6b
where we used the notation that
a box around a (sub)graph indicates that the integrand
of this (sub)graph is Taylor expanded up to the desired order
in its external momenta.
This will be the standard notation in the following.
Please note that the nullified term in figure 6b is a
massless tadpole.
Although the expansion procedure produces new ultraviolet and infrared
divergences in separate terms of the r.h.s. in figure 6b,
 these terms are well defined
for non-zero $\varepsilon$ and these new divergences cancel in the sum.   

\[ \vspace{2.2cm} \]
\[ \parbox{14cm}{ \small
{\bf Figure 6b.} The diagrammatic large mass expansion of the diagram 6a.
} \]

Let us consider another diagram presented in figure 7a.
The same reasoning as for diagram 6a holds but now we should first
add and subtract the small momentum expansion of the massive propagator 
as it is shown in the r.h.s. of figure 7a.

\[ \vspace{1.8cm} \]
\[ \parbox{14cm}{ \small {\bf Figure 7a.} A scalar triangle diagram
with one massive line  } \]

After this we can expand the expression in the square brackets
in its external momenta.
Using again the property that massless tadpoles vanish in dimensional
regularization we are left with the diagrammatic expansion
in figure 7b. Please note that the nullified term in figure 7b is a
massless tadpole.
\pagebreak

\[ \vspace{2cm} \]
\[ \parbox{14cm}{ \small {\bf Figure 7b.}
The diagrammatic large mass expansion of the diagram 7a. } \]

The triangle diagram of figure 7 can be part of a larger diagram such as
the scalar diagram presented in figure 8.
For figure 8, the same reasoning as for the diagrams in figures 6 and 7 holds 
and we are left with the indicated diagrammatic expansion.
\[ \vspace{.8cm} \]
\[ \parbox{14cm}{\small {\bf Figure 8 }
The diagrammatic large mass expansion of a 2-loop diagram. } \]

The large mass expansion of two loop propagator type diagrams is also 
treated in the recent article ref. \cite{davy0}.
It interesting to note that an analogous reasoning (i.e. focusing on
infrared regions)
can be applied to understand the related problem of
small mass expansions for which a recipe is given in 
\cite{gorishny1,dst}.   

Let us now formulate the recipe 
 for the large mass expansion of diagrams.
A line with a large mass, M, will be called a {\em heavy line}.
\\ \\
First we have to find all  asymptotically irreducible (sub)graphs.
An {\em asymptotically irreducible (sub)graph}, $g_{AI}$,
is a connected subgraph which contains at least one heavy line
(and all heavy lines connected to this one via heavy lines)
and which can not be made disconnected by cutting a non-heavy line. 
We have to expand each asymptotically irreducible (sub)graph
as a Taylor series in terms of its external momenta. 
We graphically indicate this by drawing a box around the 
asymptotically irreducible (sub)graph.
Then the large mass expansion of the whole Feynman diagram is the 
sum over all combinations of non-overlapping boxes that can be
drawn in this diagram so that all heavy lines are in boxes.
Note that a box does never cut a heavy line.
\\ \\
Or in a symbolic form:
\begin{equation}
G = \sum_{\left\{g_{AI}\right\}} G
  \hspace{-.15cm}\not
\hspace{.15cm} \mbox{\raisebox{-.1cm}{$ {\left\{g_{AI}\right\}}$ }}
 *
  \mbox{T}^{(N)} \left\{g_{AI}\right\}  + O(\frac{1}{M^{N}}),
\end{equation}
where G is the Feynman graph to be expanded. The sum goes over all
sets, $\left\{g_{AI}\right\}$, of non-overlaping asymptotically
irreducible (sub)graphs comprising all heavy lines.
$\mbox{T}^{(N)} \left\{g_{AI}\right\}$ denotes the Taylor expansion
of (sub)graphs from the set $\left\{g_{AI}\right\}$
 in their external momenta until the necessary
order. 
$ G / {\left\{g_{AI}\right\}}$
 is the graph obtained from G by 
shrinking the subgraphs from $\left\{g_{AI}\right\}$ into points.

As a practical example
we will now treat the diagrams that contribute to the axial vector part
of the Z boson decay rate.
At the 3-loop level the only Feynman diagrams that contribute 
 are
 so called 'double triangle diagrams' (the first topology in figure 3)
 with the triangles formed by bottom and top fermion loops.
This correction was originally calculated in \cite{kniehl1} and confirmed in
\cite{chetyrkin1}.
 The diagram with two  massive
triangles is zero since the only physical cut is through two gluon
propagators and $Z^{0}$ decay in two gluons is kinematically forbidden
(Landau-Yang theorem).
We are therefore left with two diagrams to be calculated,
one massive and one massless diagram. The massless diagram can be
directly calculated with the package MINCER.
The massive diagram is calculated with the above recipe.

A diagrammatic representation of the ultraviolet R operation followed by the
asymptotic expansion procedure,
applied to the massive double triangle diagram is given in figure
 \nolinebreak 9. 

\[ \vspace{4.6cm} \]
\[ \parbox{14cm}{ \small
 {\bf Figure 9.} 
Thick lines indicate top quark propagators, thin lines indicate
the massless quark propagators and spiral lines - gluon propagators.
Ultraviolet counterterms are indicated between round brackets.
The asymptotically irreducible (sub)graphs 
are surrounded by boxes with the corresponding
 tadpole topologies indicated below.
} \]
After the Taylor expansions 
the subgraphs in boxes are
reduced to massive vacuum integrals and the
resulting master topologies are indicated under the boxes. 
When the boxed subgraphs
are integrated out we are left with massless diagrams that can
be calculated with the package MINCER.
We have written efficient FORM procedures to perform the necessary massive
vacuum integrals.
The procedures include 3-loop topologies of the type Benz and non-planar
and use recursions based on the recursion scheme of \cite{broadhurst1}.
They are essential for the large mass expansion of 4-loop diagrams.
We should emphasise that only with very efficient massive vacuum 
procedures, that can deal with tensor numerators, the 4-loop calculation
is feasible. 
\newpage
Making the Taylor expansions deep enough,
we find the results for the contributions in figure 9.
Adding the massless double triangle diagram
and taking the imaginary part by applying $\ln(-s-i\epsilon) = \ln(s)
  - i\pi $
 (remember that $s = m_{Z}^2$ is the squared 4-momentum of the Z boson)
we reproduced the large mass expansion of the known result \cite{kniehl1}
\begin{equation}
\Gamma_{had}^{A,S} = \frac{G_{F}m_{Z}^{3}}{8\sqrt{2} \pi}
 \left( \frac{\alpha_{s}}{\pi} \right)^{2}
\left( d_{2}^{0} + d_{2}^{1}\frac{s}{m_{t}^2}+d_{2}^{2}\frac{s^2}{m_{t}^4}
    +d_{2}^{3}\frac{s^3}{m_{t}^6} \right)
\end{equation}
with
$ d_{2}^{0} = T_{F}^2 D \left[ - \frac{37}{24} +
 \frac{1}{2}\ln(\frac{s}{m_{t}^2}) \right]$,
$ d_{2}^{1} = T_{F}^2 D \left( \frac{7}{162}  \right)$,
$ d_{2}^{2} = T_{F}^2 D \left( \frac{7}{2400} \right)$,
$ d_{2}^{3} = T_{F}^2 D \left( \frac{52}{165375} \right)$,
$D = n^2-1$ is the number of generators 
of the colour group $SU(n)$ ($D=8$ for QCD). 

We will now show that the same method can be used to compute the $\alpha_{s}^3$
correction to $\Gamma_{had}$.  In contrast to the previous 3-loop case
where we could determine both the real and imaginary parts of a diagram,
we are now able to compute only the necessary imaginary part of diagrams with
the present available techniques (we don't have analytical results for general
4-loop massive tadpoles).
In order to find
the imaginary part of a diagram we need to calculate only terms that contain
logarithms of $s$
(since we take the imaginary part by applying 
$\ln(-s-i\epsilon) = \ln(s) - i\pi $).
 In dimensional regularization
every massless propagator type integral receives a factor
$ (\frac{\mu^{2}}{s})^{\varepsilon} = 1 + \varepsilon \ln(\frac{\mu^{2}}{s}) +
O(\varepsilon^2) $. In contrast, vacuum massive integrals produce factors
$ (\frac{\mu^{2}}{m_t^{2}})^{\varepsilon} $ ($m_t$ is the top quark mass)
 and do {\em not} give logarithms of  s.
This means that contributions without massless propagator type 
integrals do not have to be considered for our purpose.
 The expansion of one renormalized
4-loop diagram that contributes to $\Gamma_{had}^{A,S}$
is presented in figure 10.
\[ \vspace{5.2cm} \]
\[ \parbox{14cm}{\small 
{\bf Figure 10.} The large mass expansion of a renormalized
4-loop diagram} \]
In figure 10 we have only presented terms which contribute to the imaginary part
of the diagram.
All 4-loop massive diagrams contributing to $\Gamma_{had}$
give  after the large mass expansion
topologies that
can be calculated with MINCER (the massless parts) and with
the tadpole procedures (the vacuum massive parts). 
\newpage
\section{The results for $\Gamma_{had}$}
The results of the $\alpha_s^3$ approximation 
for $\Gamma_{had}^{V,NS}$, $\Gamma_{had}^{A,NS}$,
 $\Gamma_{had}^{V,S}$ and $\Gamma_{had}^{A,S}$ in the order $1/m_t^6$
of the large top quark mass expansion are presented below.
Note that these results are in 6-flavour QCD, $N_f=6$ and the
decoupling relation (\ref{decouplingrelation}) has not been applied yet.
 \[
\Gamma_{had}^{V(A),NS} =
\frac{G_{F}m_{Z}^{3}}{2\pi\sqrt{2}} \sum_{i=1}^{5}
(g_{V(A)}^i)^{2} \left( \frac{n}{3} \right)
\left[ 1+ \left( \frac{\alpha_{s}^{(6)}}{\pi} \right)  b_{1}^{0}  +
\left( \frac{\alpha_{s}^{(6)}}{\pi} \right)^{2}
\left( b_{2}^{0} + b_{2}^{1}\frac{s}{m_{t}^2}+b_{2}^{2}\frac{s^2}{m_{t}^4}
    +b_{2}^{3}\frac{s^3}{m_{t}^6} \right) \right. \hspace{5cm} \]
\begin{equation} \left. \hspace{4cm} \label{mainvns}
 + \left( \frac{\alpha_{s}^{(6)}}{\pi} \right)^{3}
\left( b_{3}^{0} + b_{3}^{1}\frac{s}{m_{t}^2}+b_{3}^{2}\frac{s^2}{m_{t}^4}
    +b_{3}^{3}\frac{s^3}{m_{t}^6} \right)
\right] ,
\end{equation}
\\
$ b_{1}^{0} = C_{F}  ( \frac{3}{4} ) $,
\\ \\
$ b_{2}^{0} =
       N_f T_F C_F  \left[  - \frac{11}{8} + \zeta_3
           + \frac{1}{4} \ln(\frac{s}{\mu^2})  \right]
       + T_F C_F  \left[ \frac{11}{8} - \zeta_3
             - \frac{1}{4} \ln(\frac{s}{m_{t}^2}) \right] $

$       + C_A C_F  \left[ \frac{123}{32} - \frac{11}{4} \zeta_3
  - \frac{11}{16} \ln(\frac{s}{\mu^2}) \right]
       + C_F^2  \left(  - \frac{3}{32} \right) ,$
\\ \\
$ b_{2}^{1} =
  T_F C_F \left[ \frac{22}{225} - \frac{1}{45} \ln(\frac{s}{m_{t}^2})  \right],
$
\\ \\
$ b_{2}^{2} =
 T_F C_F  \left[  - \frac{1303}{705600} + \frac{1}{1680}
         \ln(\frac{s}{m_{t}^2}) \right] ,$
\\ \\
$ b_{2}^{3} =
 T_F C_F  \left[ \frac{1643}{17860500} - \frac{1}{28350}
         \ln(\frac{s}{m_{t}^2})  \right], $
\\ \\
$ b_{3}^{0} =
       N_f T_F C_A C_F \left[  - \frac{485}{27}
         + \frac{112}{9}\zeta_3
         + \frac{5}{6} \zeta_5 + \frac{11}{72} \pi^2
        - \frac{11}{3} \zeta_3 \ln(\frac{s}{\mu^2})
      + \frac{259}{48} \ln(\frac{s}{\mu^2}) - \frac{11}{24} \ln^2(\frac{s}{\mu^2
})
         \right] $

$       + C_A^2 C_F \left[ \frac{90445}{3456}
      - \frac{2737}{144} \zeta_3 - \frac{55}{24} \zeta_5
    - \frac{121}{576} \pi^2
             + \frac{121}{24} \zeta_3 \ln(\frac{s}{\mu^2})
          - \frac{485}{64} \ln(\frac{s}{\mu^2}) +
       \frac{121}{192} \ln^2(\frac{s}{\mu^2}) \right] $

$      + N_f^2 T_F^2 C_F \left[ \frac{151}{54}
          - \frac{19}{9}\zeta_3
          - \frac{1}{36} \pi^2
          + \frac{2}{3} \zeta_3 \ln(\frac{s}{\mu^2})
          - \frac{11}{12} \ln(\frac{s}{\mu^2})
          + \frac{1}{12} \ln^2(\frac{s}{\mu^2}) \right] $

$      + T_F^2 C_F  \left[ \frac{151}{54}
         - \frac{19}{9} \zeta_3 - \frac{1}{36} \pi^2
         + \frac{2}{3} \zeta_3 \ln(\frac{s}{m_{t}^2})
          - \frac{11}{12}\ln(\frac{s}{m_{t}^2})
          + \frac{1}{12} \ln^2(\frac{s}{m_{t}^2}) \right] $

$      + T_F C_A C_F \left[ \frac{979}{54}
          - \frac{112}{9} \zeta_3 - \frac{5}{6}\zeta_5 - \frac{11}{72} \pi^2
        + \frac{11}{6} \zeta_3 \ln(\frac{\mu^2}{m_{t}^2}) \right. $

$  \left.\hspace{.5cm}  + \frac{11}{3} \zeta_3 \ln(\frac{s}{\mu^2})
 + \frac{11}{24} \ln(\frac{s}{m_{t}^2}) \ln(\frac{s}{\mu^2})
            - \frac{23}{8} \ln(\frac{\mu^2}{m_{t}^2})
    - \frac{259}{48} \ln(\frac{s}{\mu^2})
         \right] $

$      + N_f T_F^2 C_F \left[  - \frac{151}{27}
  + \frac{38}{9} \zeta_3 + \frac{1}{18} \pi^2
            - \frac{2}{3} \zeta_3 \ln(\frac{\mu^2}{m_{t}^2}) \right. $

$ \left.  \hspace{.5cm}   - \frac{4}{3} \zeta_3 \ln(\frac{s}{\mu^2})
    - \frac{1}{6} \ln(\frac{s}{m_{t}^2}) \ln(\frac{s}{\mu^2})
    + \frac{11}{12} \ln(\frac{\mu^2}{m_{t}^2})
          + \frac{11}{6} \ln(\frac{s}{\mu^2})
         \right] $

$     + T_F C_F^2  \left[ \frac{1}{4} - \frac{19}{4} \zeta_3 + 5 \zeta_5
         + \frac{1}{4} \ln(\frac{\mu^2}{m_{t}^2})
         - \frac{1}{8} \ln(\frac{s}{\mu^2}) \right] $

$       + C_A C_F^2  \left[  - \frac{127}{64} - \frac{143}{16} \zeta_3
  + \frac{55}{4} \zeta_5 + \frac{11}{64} \ln(\frac{s}{\mu^2}) \right] $

$     + N_f T_F C_F^2 \left[  - \frac{29}{64} + \frac{19}{4}\zeta_3
          - 5 \zeta_5 + \frac{1}{8} \ln(\frac{s}{\mu^2}) \right]
    + C_F^3  \left(  - \frac{69}{128} \right) $ ,
\newpage
\noindent$ b_3^1 =
      T_F C_F^2 \left[ \frac{93599}{364500} - \frac{529}{2160} \zeta_3
           + \frac{11}{1620} \pi^2
          - \frac{1}{135} \ln(\frac{\mu^2}{m_{t}^2}) \ln(\frac{s}{\mu^2})
    \right.$

$ \left.\hspace{.5cm}
          - \frac{1333}{24300} \ln(\frac{\mu^2}{m_{t}^2})
          + \frac{7}{540} \ln^2(\frac{\mu^2}{m_{t}^2})
          + \frac{1421}{24300}  \ln(\frac{s}{\mu^2})
          - \frac{11}{540}  \ln^2(\frac{s}{\mu^2}) \right]  $

$  +T_F C_A C_F  \left[  - \frac{716839}{5832000}
      + \frac{53}{864} \zeta_3 - \frac{11}{3240} \pi^2
        - \frac{11}{540} \ln(\frac{\mu^2}{m_{t}^2}) \ln(\frac{s}{\mu^2})
     \right.$

$ \left.\hspace{.5cm}
        + \frac{27547}{194400} \ln(\frac{\mu^2}{m_{t}^2})
          - \frac{11}{360} \ln^2(\frac{\mu^2}{m_{t}^2})
           - \frac{7301}{194400} \ln(\frac{s}{\mu^2})
          + \frac{11}{1080} \ln^2(\frac{s}{\mu^2})  \right]  $

$+ T_F^2 C_F \left[ \frac{8927}{121500} - \frac{1}{405} \pi^2
         - \frac{7}{135} \ln(\frac{s}{m_{t}^2})
         + \frac{1}{135} \ln^2(\frac{s}{m_{t}^2})  \right] $

$+N_f T_F^2 C_F  \left[  - \frac{1513}{30375} + \frac{1}{405} \pi^2
                - \frac{1}{75} \ln(\frac{\mu^2}{m_{t}^2})
          + \frac{1}{135} \ln^2(\frac{\mu^2}{m_{t}^2})
     \right.$

$ \left.\hspace{.5cm}
          + \frac{7}{135} \ln(\frac{s}{\mu^2})
          - \frac{1}{135} \ln^2(\frac{s}{\mu^2}) \right] , $
\\ \\
\noindent$ b_3^2 =
         T_F C_F^2 \left[  - \frac{211867}{7112448}
          + \frac{2291}{322560} \zeta_3
          + \frac{43}{120960} \pi^2
           - \frac{79}{20160} \ln(\frac{\mu^2}{m_{t}^2}) \ln(\frac{s}{\mu^2})
     \right.$

$ \left.\hspace{.5cm}
           + \frac{37109}{2540160} \ln(\frac{\mu^2}{m_{t}^2})
          - \frac{23}{8064} \ln^2(\frac{\mu^2}{m_{t}^2})
           + \frac{126523}{12700800} \ln(\frac{s}{\mu^2})
          - \frac{43}{40320} \ln^2(\frac{s}{\mu^2}) \right] $

$     + T_F C_A C_F \left[  - \frac{61445851}{5334336000}
           + \frac{271}{18432} \zeta_3 - \frac{43}{241920} \pi^2
         + \frac{29}{13440} \ln(\frac{\mu^2}{m_{t}^2}) \ln(\frac{s}{\mu^2})
     \right.$

$ \left.\hspace{.5cm}
         - \frac{347593}{50803200} \ln(\frac{\mu^2}{m_{t}^2})
         + \frac{131}{80640} \ln^2(\frac{\mu^2}{m_{t}^2})
         - \frac{175597}{50803200} \ln(\frac{s}{\mu^2})
         + \frac{43}{80640} \ln^2(\frac{s}{\mu^2}) \right] $

$   + T_F^2 C_F \left[ \frac{175039}{222264000}
           + \frac{1}{15120} \pi^2
           + \frac{109}{151200} \ln(\frac{s}{m_{t}^2})
           - \frac{1}{5040} \ln^2(\frac{s}{m_{t}^2}) \right] $

$   + N_f T_F^2 C_F \left[ \frac{15209}{111132000}
           - \frac{1}{15120} \pi^2
          + \frac{37}{264600} \ln(\frac{\mu^2}{m_{t}^2})
          - \frac{1}{5040} \ln^2(\frac{\mu^2}{m_{t}^2})
    \right.$

$ \left.\hspace{.5cm}
          - \frac{11}{10080} \ln(\frac{s}{\mu^2})
          + \frac{1}{5040} \ln^2(\frac{s}{\mu^2})  \right] , $
\\ \\
 \noindent$ b_3^3 =
     T_F C_F^2 \left[ \frac{3395030363}{5401015200000}
          - \frac{121}{268800} \zeta_3 + \frac{1}{5103000} \pi^2
         + \frac{67}{425250} \ln(\frac{\mu^2}{m_{t}^2}) \ln(\frac{s}{\mu^2})
     \right.$

$ \left.\hspace{.5cm}
        - \frac{31991}{61236000} \ln(\frac{\mu^2}{m_{t}^2})
        + \frac{269}{1701000} \ln^2(\frac{\mu^2}{m_{t}^2})
         - \frac{69173}{428652000} \ln(\frac{s}{\mu^2})
         - \frac{1}{1701000} \ln^2(\frac{s}{\mu^2}) \right] $

$    + T_F C_A C_F \left[ \frac{22635964507}{2057529600000}
                - \frac{6217}{691200} \zeta_3 - \frac{1}{10206000} \pi^2
          - \frac{109}{1701000} \ln(\frac{\mu^2}{m_{t}^2}) \ln(\frac{s}{\mu^2})
     \right.$

$ \left.\hspace{.5cm}
          + \frac{1937441}{11430720000} \ln(\frac{\mu^2}{m_{t}^2})
          - \frac{73}{1134000} \ln^2(\frac{\mu^2}{m_{t}^2})
          + \frac{28963}{34292160000} \ln(\frac{s}{\mu^2})
          + \frac{1}{3402000} \ln^2(\frac{s}{\mu^2})  \right] $

$       + T_F^2 C_F \left[  - \frac{5423329}{67512690000}
              - \frac{1}{255150} \pi^2
          - \frac{11}{425250} \ln(\frac{s}{m_t^2})
          + \frac{1}{85050} \ln^2(\frac{s}{m_t^2})\right] $

$      + N_f T_F^2 C_F \left[ \frac{81169}{16878172500}
          + \frac{1}{255150} \pi^2
          - \frac{199}{53581500} \ln(\frac{\mu^2}{m_{t}^2})
          + \frac{1}{85050} \ln^2(\frac{\mu^2}{m_{t}^2})
     \right.$

$ \left.\hspace{.5cm}
          + \frac{7}{121500} \ln(\frac{s}{\mu^2})
          - \frac{1}{85050} \ln^2(\frac{s}{\mu^2})  \right], $
\\ \\
where $n$ is the parameter of the colour group $SU(n)$
($n=3$ for QCD).
Note that the coefficient $b_{2}^{1}$ agrees with ref. \cite{che}
and $b_{2}^{2}$, $b_{2}^{3}$ agree with the expansion of the exact
result ref. \cite{kniehl2}.

\newpage
\[
\Gamma_{had}^{A,S} = \frac{G_{F}m_{Z}^{3}}{8\sqrt{2} \pi}
\left[ \left( \frac{\alpha_{s}^{(6)}}{\pi} \right)^{2}
\left( d_{2}^{0} + d_{2}^{1}\frac{s}{m_{t}^2}+d_{2}^{2}\frac{s^2}{m_{t}^4}
    +d_{2}^{3}\frac{s^3}{m_{t}^6} \right) \right. \hspace{6cm} \]
\begin{equation} \label{mainas} \hspace{5cm}
\left.  + \left( \frac{\alpha_{s}^{(6)}}{\pi} \right)^{3}
\left( d_{3}^{0} + d_{3}^{1}\frac{s}{m_{t}^2}+d_{3}^{2}\frac{s^2}{m_{t}^4}
    +d_{3}^{3}\frac{s^3}{m_{t}^6} \right)
\right] ,
\end{equation}
\\
$ d_{2}^{0} = T_{F}^2 D \left[ - \frac{37}{24} +
 \frac{1}{2}\ln(\frac{s}{m_{t}^2}) \right]$,
 \\ \\
$ d_{2}^{1} = T_{F}^2 D \left( \frac{7}{162}  \right)$,
 \\ \\
$ d_{2}^{2} = T_{F}^2 D \left( \frac{7}{2400} \right)$,
 \\ \\
$ d_{2}^{3} = T_{F}^2 D \left( \frac{52}{165375} \right)$,
 \\ \\
$ d_{3}^{0} = N_{f} T_{F}^3 D \left[ \frac{25}{36} - \frac{1}{18}\pi^2 +
 \frac{1}{9} \ln(\frac{\mu^2}{m_{t}^2})
 - \frac{1}{6} \ln^{2}(\frac{\mu^2}{m_{t}^2}) - \frac{11}{12}
 \ln(\frac{s}{\mu^2}) + \frac{1}{6} \ln^{2}(\frac{s}{\mu^2})
 \right] $

\hspace{.0cm}$+ C_{A}T_{F}^2 D \left[  - \frac{215}{48} - \frac{1}{2}\zeta_{3}
 + \frac{11}{72} \pi^2
+ \frac{19}{36} \ln(\frac{\mu^2}{m_{t}^2})
+ \frac{11}{24} \ln^2 (\frac{\mu^2}{m_{t}^2})
+ \frac{161}{48} \ln(\frac{s}{\mu^2})
- \frac{11}{24} \ln^2(\frac{s}{\mu^2}) \right]$

\hspace{.0cm}$ + C_{F}T_{F}^2 D \left[ - \frac{3}{4} + \frac{3}{2} \zeta_3
 - \frac{3}{4} \ln(\frac{\mu^2}{m_{t}^2}) \right] +
 T_{F}^3 D \left[ - \frac{157}{108} +\frac{1}{18}\pi^{2}
+ \frac{11}{12} \ln(\frac{s}{m_{t}^2})
- \frac{1}{6} \ln^{2} (\frac{s}{m_{t}^2}) \right], $
\\ \\
$ d_{3}^{1} =
       N_{f} T_{F}^3 D \left[ \frac{181}{3888}
          - \frac{17}{486} \ln(\frac{\mu^2}{m_{t}^2})
          - \frac{1}{162} \ln(\frac{s}{\mu^2}) \right]$

$   + C_{A}T_{F}^2 D \left[  - \frac{3983}{31104} + \frac{13}{256}\zeta_3 +
      \frac{187}{1944} \ln(\frac{\mu^2}{m_{t}^2})
      + \frac{11}{648} \ln(\frac{s}{\mu^2}) \right] $

$   + C_{F}T_{F}^2 D \left[  - \frac{1109}{5184} + \frac{235}{1152}\zeta_3
            - \frac{7}{108} \ln(\frac{\mu^2}{m_{t}^2}) \right] $

$       + T_{F}^3 D \left[ - \frac{293039}{1458000} + \frac{7}{108}\zeta_3
        + \frac{1}{270} \pi^2
         + \frac{128}{2025} \ln(\frac{s}{m_{t}^2})
       - \frac{1}{90} \ln^{2}(\frac{s}{m_{t}^2}) \right] ,$
\\ \\
$ d_{3}^{2} =
N_{f} T_{F}^3 D \left[ \frac{14321}{5832000}
            - \frac{259}{97200}\ln(\frac{\mu^2}{m_{t}^2})
          - \frac{7}{9720} \ln(\frac{s}{\mu^2}) \right]$

$       + C_{A}T_{F}^2 D \left[ - \frac{54654577}{746496000}
         + \frac{390451}{6635520}\zeta_3
          + \frac{2849}{388800} \ln(\frac{\mu^2}{m_{t}^2})
         + \frac{77}{38880} \ln(\frac{s}{\mu^2}) \right] $

$      + C_{F}T_{F}^2 D \left[  - \frac{12499297}{24883200} +
             \frac{463087}{1105920}\zeta_3
          - \frac{7}{800} \ln(\frac{\mu^2}{m_{t}^2}) \right] $

$       + T_{F}^3 D \left[ - \frac{1672471517}{64012032000} +
          \frac{1519}{110592}\zeta_3 + \frac{1}{2520}\pi^2
          + \frac{13163}{2381400} \ln(\frac{s}{m_{t}^2})
          - \frac{1}{840} \ln^{2}(\frac{s}{m_{t}^2}) \right] ,$
\\ \\
$ d_{3}^{3} =
       N_{f} T_{F}^3 D \left[ \frac{2588111}{10001880000}
          - \frac{4973}{15876000} \ln(\frac{\mu^2}{m_{t}^2})
          - \frac{47}{453600} \ln(\frac{s}{\mu^2}) \right] $

$       + C_{A}T_{F}^2 D \left[  - \frac{279724766851}{4096770048000} +
         \frac{50083309}{884736000}\zeta_3
        + \frac{54703}{63504000} \ln(\frac{\mu^2}{m_{t}^2})
        + \frac{517}{1814400} \ln(\frac{s}{\mu^2}) \right] $

$       + C_{F}T_{F}^2 D \left[ - \frac{423770834093}{487710720000} +
         \frac{9946661}{13762560} \zeta_3
          - \frac{26}{18375} \ln(\frac{\mu^2}{m_{t}^2}) \right] $

$       + T_{F}^3 D \left[ - \frac{17132293649}{3292047360000} +
           \frac{4823}{1474560}\zeta_3 + \frac{1}{17010}\pi^2
          + \frac{21353}{28576800} \ln(\frac{s}{m_{t}^2})
          - \frac{1}{5670} \ln^{2}(\frac{s}{m_{t}^2}) \right].$

\newpage
For completeness we also present  separately the bottom-bottom
and top-top 
contributions to $\Gamma_{had}^{A,S}$. ($\Gamma_{had}^{A,S} =
 \Gamma_{had}^{A,S,bb} + \Gamma_{had}^{A,S,tt} + \Gamma_{had}^{A,S,tb}$
 is the sum over three possible pairs of the weak coupling constants:
 the bottom-bottom, top-top and top-bottom contributions.)
\[
\Gamma_{had}^{A,S,bb}
 = \frac{G_{F}m_{Z}^{3}}{2\sqrt{2} \pi}
\left( g_A^{bot}\right)^2
\left[ \left( \frac{\alpha_{s}^{(6)}}{\pi} \right)^{2}
 d_{2}^{0,b}  \right. \hspace{9cm} \]
\begin{equation} \left. \hspace{2cm}
 + \left( \frac{\alpha_{s}^{(6)}}{\pi} \right)^{3}
\left( d_{3}^{0,b} + d_{3}^{1,b}\frac{s}{m_{t}^2}+d_{3}^{2,b}\frac{s^2}{m_{t}^4}
    +d_{3}^{3,b}\frac{s^3}{m_{t}^6} \right)
\right] ,
\end{equation}
\\
$ d_{2}^{0,b} = T_{F}^2 D \left[ - \frac{17}{12} +
 \frac{1}{2}\ln(\frac{s}{\mu^2}) \right]$,
 \\ \\
$ d_{3}^{0,b} = C_{F}T_{F}^2 D \left[
  - \frac{53}{96} + \frac{1}{2} \zeta_3 \right] $

$+ N_{f} T_{F}^3 D \left[
  \frac{923}{648} - \frac{1}{18}\pi^2 - \frac{11}{12} \ln(\frac{s}{\mu^2})
         + \frac{1}{6} \ln^2(\frac{s}{\mu^2}) 
 \right] $

\hspace{.0cm}$+ C_{A}T_{F}^2 D \left[
  - \frac{16231}{2592} + \frac{17}{24} \zeta_3 + \frac{11}{72} \pi^2 
    + \frac{161}{48} \ln(\frac{s}{\mu^2}) 
     - \frac{11}{24} \ln^2(\frac{s}{\mu^2}) 
 \right]$

$
       + T_{F}^3 D \left[
 - \frac{211}{108} + \frac{1}{18} \pi^2 + \frac{11}{12} \ln(\frac{s}{m_{t}^2})
         - \frac{1}{6} \ln^2(\frac{s}{m_{t}^2}) 
\right], $
\\ \\
$ d_{3}^{1,b} =
        T_{F}^3 D \left[
 - \frac{134}{1125} + \frac{1}{270} \pi^2 
  + \frac{77}{1350} \ln(\frac{s}{m_{t}^2})
       - \frac{1}{90} \ln^2(\frac{s}{m_{t}^2})
 \right] ,$
\\ \\
$ d_{3}^{2,b} =
        T_{F}^3 D \left[ 
      - \frac{673639}{74088000} + \frac{1}{2520} \pi^2
      + \frac{53}{11025} \ln(\frac{s}{m_{t}^2})
      - \frac{1}{840} \ln^2(\frac{s}{m_{t}^2})
 \right] ,$
\\ \\
$ d_{3}^{3,b} =
        T_{F}^3 D \left[
    - \frac{2679041}{2250423000} + \frac{1}{17010} \pi^2
      + \frac{2299}{3572100} \ln(\frac{s}{m_{t}^2}) 
          - \frac{1}{5670} \ln^2(\frac{s}{m_{t}^2})
\right].$
\\ \\
Note that 
except for the massless diagrams there is only one masssive diagram 
that contributes to $\Gamma_{had}^{A,S,bb}$ (the $3^{rd}$ topology
in the $3^{rd}$ line of figure 3).

\begin{equation}
\Gamma_{had}^{A,S,tt}
 = \frac{G_{F}m_{Z}^{3}}{2\sqrt{2} \pi}
\left( g_A^{top}\right)^2
  \left( \frac{\alpha_{s}^{(6)}}{\pi} \right)^{3}
\left( d_{3}^{0,t} + d_{3}^{1,t}\frac{s}{m_{t}^2}+d_{3}^{2,t}\frac{s^2}{m_{t}^4}
    +d_{3}^{3,t}\frac{s^3}{m_{t}^6} \right) ,
\end{equation}
$ d_{3}^{0,t} = d_{3}^{1,t} = 0,$
\\ \\
$ d_{3}^{2,t} =
        T_{F}^3 D (N_f-1) \left( \frac{1}{8640} \right) ,$
\\ \\
$ d_{3}^{3,t} =
        T_{F}^3 D (N_f-1) \left( \frac{13}{583200} \right). $
\\ \\
Note also that although separate diagrams that contribute to
 $\Gamma_{had}^{A,S,tt}$  are generally
non-zero, they cancel in the sum except for the one diagram
(the $3^{rd}$ topology in the $3^{rd}$ line of figure 3)
that has a colour factor proportional to $T_{F}^3 D (N_f-1)$.
This fact can be understood through the operator product expansion
technique ref. \cite{gorishny3}.
In addition note that the imaginary part of the 
diagram with three top quark loops 
(the $3^{rd}$ topology in the $3^{rd}$ line of figure 3) vanishes
according to the Landau-Yang theorem (which we checked explicitly).

The result for the vector singlet part reads
 \[
\Gamma_{had}^{V,S} =
\frac{G_{F}m_{Z}^{3}}{2\pi\sqrt{2}} \left( \sum_{i=1}^{5}
g_{V}^i \right)^2
\left( \frac{\alpha_{s}^{(6)}}{\pi} \right)^{3} c_{3}^{0} \hspace{5cm} \]
\begin{equation} \label{mainvs}
 + \frac{G_{F}m_{Z}^{3}}{2\pi\sqrt{2}}  g_{V}^{top}
\left(  \sum_{i=1}^{5} g_{V}^i \right)
  \left( \frac{\alpha_{s}^{(6)}}{\pi} \right)^{3}
\left(  c_{3}^{1}\frac{s}{m_{t}^2}+c_{3}^{2}\frac{s^2}{m_{t}^4}
    +c_{3}^{3}\frac{s^3}{m_{t}^6} \right),
\end{equation}
\\ \\
$ c_{3}^{0} = T_F^2 d^{abc} d_{abc} \left(
  \frac{11}{144} - \frac{1}{6} \zeta_3 \right) $
\\ \\
$ c_{3}^{1} =
 T_F^2 d^{abc} d_{abc} \left(  - \frac{37}{576} + \frac{13}{216}
              \zeta_3 \right)  $
\\ \\
$ c_{3}^{2} = T_F^2 d^{abc} d_{abc} \left(  - \frac{110401}{2332800}
       + \frac{29}{720} \zeta_3 \right)  $
\\ \\
$ c_{3}^{3} =
        T_F^2 d^{abc} d_{abc} \left(  - \frac{45317647}{1143072000} +
         \frac{5009}{151200}\zeta_3 \right) , $
\\ \\
where $d^{abc}$ are the symmetrical structure constants of 
the $SU(n)$ colour group, $d^{abc} d_{abc}= 40/3$ for QCD.
Note that the constant $ c_{3}^{0}$ comes from massless diagrams only
(although separate massive diagrams are non-zero in the leading order
of the large top mass expansion, they add up to zero)
as it is required by the decoupling mechanism for the vector part.
It is interesting to
mention that the diagrams with two massive top quark loops are
non-zero separately but they add up to zero in all orders of
the large mass expansion.

Explicit checks show that the coefficients of the logarithms 
in eqs. (\ref{mainvns}) and (\ref{mainas}) 
are in agreement with the required
renormalization group invariance of the physical quantity which in 
the $\alpha_{s}^3$ approximation reads
\begin{equation} \mu^{2} \frac{d}{d \mu^{2}} \Gamma
 (\mu^{2},\alpha_{s}(\mu),m_{t} (\mu) ) = O\left(\alpha_{s}^4\right).
\end{equation}
Of course the true physical quantity is $\Gamma_{had}= 
\Gamma_{had}^{V,NS}+\Gamma_{had}^{V,S}
+\Gamma_{had}^{A,NS}+\Gamma_{had}^{A,S}$.
 But from a theoretical point
of view each of the four separate parts is renormalized independently
and is therefore renormalization group invariant by itself.

The results that are presented in this section were obtained in an arbitrary
covariant gauge for the gluon fields i.e. keeping the gauge parameter as a
free parameter in the calculations.
The explicit cancellation of the gauge dependence
in the physical quantities gives a good check of the results.
Individual diagrams contain $\zeta_2$, and $\zeta_4$ but these contributions
add up to zero in the total results.

\section{Final results (in a decoupled form)}

It is known that the Appelquist-Carazzone theorem \cite{appelquist}
about the decoupling of a heavy particle in quantum field theory does not
work in its naive form for the $\overline{MS}$ renormalization scheme
and one should make an extra shift in the coupling constant 
(see eq.(\ref{decouplingrelation}) ) to  make
the decoupling explicit (i.e. to kill the large-mass logarithms)
 \cite{ovrut,bernreuther}. 
However in the presence of an axial vector current  the decoupling does not
work (even after the shift in the coupling constant) 
due to the presence of axial anomaly type diagrams as is the case for
the axial vector singlet contribution to $\Gamma_{had}$.

Since we have explicitly calculated the top quark mass terms
in the order $\alpha_s^3$ for $\Gamma_{had}$, 
we can derive the decoupling relation for the QCD coupling constant
in the next-to-next-to-leading (NNL) order.
Because the vector contribution to $\Gamma_{had}$ should
obey the decoupling mechanism, the use of the decoupling relation
should convert the new 6 flavour result for $\Gamma_{had}^{V,NS}$
(see eq.(\ref{mainvns})) to the previously known result in effective
massless 5 flavour QCD (see eq.(\ref{gammavec})). One can see that
the NNL order decoupling relation obtained in this way
 (see eq.(\ref{decouplingrelation})) slightly differs from the one
known in literature \cite{bernreuther}

In order to settle this discrepancy, we performed an independent calculation.
We have calculated the 3-loop massless quark propagator with a
zero momentum $\overline{\psi} \psi$ operator insertion
(where $\psi$ is the quark field)
\begin{equation}
 G_{\overline{\psi}[\overline{\psi}\psi ] \psi}(Q^2) =
 \int e^{iq\cdot x}  dx dy 
 <0| T \left\{ \overline{\psi_{\alpha}}(x)\overline{\psi_{\beta}}(y)
  \psi_{\beta}(y) \psi_{\alpha}(0) e^{i S} \right\} |0>
\end{equation}
We will derive  the decoupling relations 
from this (gauge dependent) Green function.
One can also use the normal quark propagator for this purpose but one
has to evaluate it at 4-loop level  to derive the decoupling relation
for the coupling constant in the NNL order. This is because at the one loop
level the quark propagator is proportional to the gauge parameter.

For a physical quantity the decoupling mechanism consists of a
shift in the coupling constants (and in the light masses, if present).
The decoupling mechanism for the Green function is more complicated 
than that for a physical quantity because the renormalization of
the Green function involves an overall renormalization constant,
$G_{ren} = Z G_{Bare}$, and one should also obtain the decoupling
relation for Z. To avoid this small complication, we prefer to 
take the quantity
\[ \frac{d}{d \ln (Q^2)} \ln
 \left( G_{\overline{\psi}[\overline{\psi}\psi ] \psi ,ren}(Q^2) \right) \]
that imitates a physical quantity in the sense that it has no 
overall renormalization constant Z. The result of our calculation in the
leading order of the large top quark mass expansion reads

\[ \frac{d}{d \ln (Q^2)} \ln
 \left( G_{\overline{\psi}[\overline{\psi}\psi ] \psi ,ren}(Q^2) \right) =
\left( \frac{\alpha_{s}^{(6)}}{\pi} \right) \left(
  C_f (1-\frac{1}{3}\xi) \right) \hspace{7cm} \]
\[ + \left( \frac{\alpha_{s}^{(6)}}{\pi} \right)^2 \left(
N_f T_F C_F \left[ -\frac{5}{9} + \frac{1}{3} \ln(\frac{Q^2}{\mu^2}) \right]
+ T_F C_F \left[ -\frac{1}{3} \ln(\frac{\mu^2}{m_t^2}) \right] \right.
 \hspace{4cm} \]
\[ \left. \hspace{1cm}
+ C_A C_F \left\{ \frac{295}{144} -\frac{11}{12}  \ln(\frac{Q^2}{\mu^2})
     + \xi \left[ -\frac{1}{6} + \frac{1}{8}  \ln(\frac{Q^2}{\mu^2}) \right]
     + \xi^2 \left[ -\frac{1}{48} + \frac{1}{24} \ln(\frac{Q^2}{\mu^2}) \right] 
 \right\} 
+ C_F^2 \frac{1}{4} \right) \]
\[ + \left( \frac{\alpha_{s}^{(6)}}{\pi} \right)^3 \left(
 N_f T_F^2 C_F \left[ - \frac{2}{9}
    \ln(\frac{\mu^2}{m_t^2}) \ln(\frac{Q^2}{\mu^2})
        + \frac{10}{27} \ln(\frac{\mu^2}{m_t^2})  \right] \right.\hspace{5cm} \]
\[
+T_F C_A C_F \left\{ \frac{2}{9} +
       \frac{11}{18} \ln(\frac{\mu^2}{m_t^2}) \ln(\frac{Q^2}{\mu^2})
     -\frac{385}{216} \ln(\frac{\mu^2}{m_t^2})  \right. \hspace{5cm}
\] 
\[ 
   + \xi \left. \left[ -\frac{89}{1728}
      -\frac{1}{24} \ln(\frac{\mu^2}{m_t^2}) \ln(\frac{Q^2}{\mu^2})
    + \frac{13}{144} \ln(\frac{\mu^2}{m_t^2})
    - \frac{1}{48}  \ln^2(\frac{\mu^2}{m_t^2}) \right]   \right\}
   \hspace{2.5cm}\]
\begin{equation} \label{psipsibar} \left.
+ T_F C_F^2 \left[ - \frac{13}{48}
          + \frac{1}{12} \ln(\frac{\mu^2}{m_t^2}) \right]
+ T_F^2 C_F \frac{1}{9} \ln^2(\frac{\mu^2}{m_t^2}) \right) +
  \left( \frac{\alpha_{s}^{(6)}}{\pi} \right)^3 \left(
  \begin{array}{c} \mbox{massless} \\ 
  \mbox{contrib.} \end{array} 
 \right),
\end{equation} 
up to an overall normalization. $\xi$  is the gauge parameter
that appears in the gluon propagator as
$ \frac{i}{q^2+i\epsilon}(-g^{\mu,\nu}+(1-\xi)
 \frac{q^{\mu}q^{\nu}}{q^2+i\epsilon})$.
For the $\alpha_s^3$ term only contributions that contain one or more
top quark loops are presented. These massive contributions of the order
 $\alpha_s^3$ 
should disappear after the application of the decoupling relations for
the coupling constant and for the gauge parameter $\xi$. 
From this requirement we find the decoupling relations in the NNL order
for the coupling constant presented in eq.(\ref{decouplingrelation}) and
for the gauge parameter presented below.
\[
 \xi^{(6)}(\mu) = \xi^{(5)}(\mu) \left\{ 1
  - \left(\frac{\alpha_{s}^{(5)}(\mu)}{\pi}\right) \frac{T_{F}}{3}
     \ln(\frac{\mu^2}{m_{t}^{2}(\mu)}) 
   +\left(\frac{\alpha_{s}^{(5)}(\mu)}{\pi}\right)^2
   \left[ -\frac{1}{16} T_F C_A \ln^2(\frac{\mu^2}{m_{t}^{2}(\mu)}) 
\right. \right. \]
\begin{equation} \label{gaugede} \hspace{2cm} \left. \left.
 + (\frac{1}{4} T_F C_F - \frac{5}{16} T_F C_A ) 
                    \ln(\frac{\mu^2}{m_{t}^{2}(\mu)}) 
  + \frac{13}{192} T_F C_A - \frac{13}{48} T_F C_F \right] 
+ O(\alpha_{s}^{3}) \right\} . \end{equation}
 Please note that the
 term $\frac{13}{48} T_F C_F$ that we found in eq.(\ref{gaugede}) 
for the gauge parameter (as well as the analogous term for the coupling
constant) is slightly different from the one in ref. \cite{bernreuther}.
Note that the decoupling relation for the coupling constant,
 eq.(\ref{decouplingrelation}), is derived by us in two
independent ways: from the 4-loop calculation of $\Gamma_{had}$ and  
the 3-loop calculation of $G_{\overline{\psi}[\overline{\psi}\psi ] \psi}$.

We will now give the results for $\Gamma_{had}$ in effective QCD
with 5 active massless flavours at the renormalization scale $\mu = m_Z$.
These results are obtained by substitution of the decoupling
relation, eq.(\ref{decouplingrelation}), into the results for 6 flavour QCD 
of eqs.(\ref{mainvns}), (\ref{mainas}) and (\ref{mainvs}).
We will use the notation $x\equiv \frac{m_Z^2}{m_t^2} $.
\[ \Gamma_{had} = \Gamma_{had}^{V,NS}+\Gamma_{had}^{A,NS}+
            \Gamma_{had}^{V,S} + \Gamma_{had}^{A,S}.\hspace{8cm} \]
\begin{equation}
\Gamma_{had}^{V(A),NS} =
\frac{G_{F}m_{Z}^{3}}{2\pi\sqrt{2}} \sum_{i=1}^{5}
(g_{V(A)}^i)^{2}
\left[ 1+ \left( \frac{\alpha_{s}^{(5)}}{\pi} \right)  b_{1}  +
\left( \frac{\alpha_{s}^{(5)}}{\pi} \right)^{2}
 b_{2} + \left( \frac{\alpha_{s}^{(5)}}{\pi} \right)^{3}  b_{3} 
   + O \left( \alpha_{s}^4\right) \right] ,
 \hspace{3cm} \end{equation}
$ b_{1}= 1$,
\\ \\
$ b_{2} =
   1.4092 $

$+  \left[  0.065185   
   - 0.014815 \ln(x)  \right] x  $

$+ \left[  - 0.0012311 
   + 0.00039683 \ln(x) \right] x^2 $

$+  \left[  0.000061327 
    - 0.000023516 \ln(x) \right] x^3 + O(x^4) ,$
\\ \\
$ b_{3} =
    - 12.767 $

$ +  \left[   - 0.17374 
    + 0.21242 \ln(x) 
    - 0.037243 \ln^2(x) \right] x $

$+   \left[   - 0.0075218 
    - 0.00058859 \ln(x) 
    + 0.00038305 \ln^2(x) \right] x^2 $

$+  \left[    0.00050411 
     - 0.00012099 \ln(x) 
     + 0.000031419 \ln^2(x) \right] x^3 + O(x^4).$
\\ \\ 
\begin{equation} 
\Gamma_{had}^{V,S} =
\frac{G_{F}m_{Z}^{3}}{2\pi\sqrt{2}} \left[ \left( \sum_{i=1}^{5}
g_{V}^i \right)^2
\left( \frac{\alpha_{s}^{(5)}}{\pi} \right)^{3} c_{3}
 +  g_{V}^{top}
\left(  \sum_{i=1}^{5} g_{V}^i \right)
  \left( \frac{\alpha_{s}^{(5)}}{\pi} \right)^{3}
 c_{3}^{top} + O \left( \alpha_{s}^4\right) \right] , \hspace{5cm} 
\end{equation}
$c_{3} = -0.41318 ,$
\\ \\
$c_{3}^{top} =
    0.027033  x 
  + 0.0036355  x^2
  + 0.00058874  x^3 + O(x^4).$
\\ \\  
\begin{equation}
\Gamma_{had}^{A,S} = \frac{G_{F}m_{Z}^{3}}{2  \pi \sqrt{2}}
\left( \frac{1}{4} \right)
\left[ \left( \frac{\alpha_{s}^{(5)}}{\pi} \right)^{2} d_{2}
+  \left( \frac{\alpha_{s}^{(5)}}{\pi} \right)^{3} d_{3} 
   + O \left( \alpha_{s}^4\right) \right] ,
 \hspace{8.5cm} 
\end{equation}
$ d_{2} =
    - 3.0833  + \ln(x) 
 + 0.086420  x
 + 0.0058333 x^2 
 + 0.00062887 x^3 + O(x^4),$
\\ \\
$ d_{3} =
 - 18.654  
 + 1.7222 \ln(x)    
   +  1.9167 \ln^2(x) $

$ + \left[   - 0.12585 
   + 0.28646  \ln(x)  
   - 0.011111 \ln^2(x)  \right] x $

$+ \left[  - 0.0031322 
      + 0.012117 \ln(x)   
    - 0.0011905 \ln^2(x)   \right] x^2 $

$+ \left[ - 0.00088827 
    + 0.00047262  \ln(x)
       - 0.00017637 \ln^2(x)   \right] x^3 + O(x^4).$

We want to stress that the results show an excellent convergence of
the large top quark mass expansion which can be seen from the 
fast decrease of the coefficients with the order of the expansion
parameter $x= \frac{m_Z^2}{m_t^2}$.

Note that the massive logarithms $\ln(\frac{m_Z^2}{m_{t}^2})$
are present in the leading order of the large mass expansion 
in $\Gamma_{had}^{A,S}$ only (the violation of
decoupling of the top quark). These logarithms can
be summed up using the operator product expansion technique
to produce a result that is finite in the limit of an infinitely large
top quark mass as it was done in ref. \cite{chetyrkin1} for the order
$\alpha_s^2$. However, for realistic values of $m_t$ the above expressions can
be trusted and are quite stable with respect to a change in the renormalization
parameter $\mu$ around the natural scale for this process $\mu = m_Z$,
as for example can be seen for $\Gamma_{had}^{A,S}$ in figure 11.
\[ \vspace{2cm} \]



\vspace{5.5cm} \[ \parbox{14cm}{ \small
{\bf Figure 11.}
The $\mu$ dependence of $\Gamma_{had}^{A,S}$ around $\mu=m_Z$ for
$m_t(m_Z)$ = 140 GeV and $\Lambda_{QCD}^{(5)}$ = 0.2 GeV.
 The dotted line indicates the 
3-loop result, the solid line indicates the result up to (and including)
4-loops.
} \]
One can use $\Gamma_{had}^{V,NS}$ from the present
paper to
obtain the hadronic decay width $\Gamma(W^{\pm} \rightarrow hadrons)$
(by a standard change of the weak coupling constants)
and to obtain the large mass expansion in the $\alpha_{s}^3$ order
for $\Gamma(\tau^{-} \rightarrow \nu_{\tau}+hadrons)$ (as we
do it in the next section).
From the results of this paper
one can also straightforwardly obtain the $\alpha_s^3$ approximation
to the total cross section of electron-positron annihilation
$\sigma_{tot}(e^+e^- \rightarrow \gamma, Z^0 \rightarrow hadrons)$
in the necessary energy range, e.g. below, or above the $Z^0$ peak.

\section{The large charm quark mass expansion for $\tau$ lepton decay}

Another prominent process (beside Z boson decay) to extract the value
of $\alpha_s$ from experiment is semihadronic $\tau$ lepton decay.
Although the scale of this process is relatively low, non-perturbative
corrections turn out to be small and 
perturbative QCD can be used \cite{schilcher}-\cite{braaten2}
 to calculate the $\tau$ lepton decay rate, or the ratio
\begin{equation} R_{\tau}= 
\frac{\Gamma (\tau^{-} \rightarrow \nu_{\tau}+hadrons) }{
  \Gamma ( \tau^{-} \rightarrow  \nu_{\tau} e^{-} \overline{\nu}_{e}) } 
\end{equation}
The ratio $R_{\tau}$ is expressed through the imaginary part of the
W boson current correlator as (modulo a contribution 
 that vanishes for massless active flavours)
\begin{equation} \label{rtau}
 R_{\tau}= 12\pi \int_{0}^{m_{\tau}^2}
\frac{ ds}{m_{\tau}^2} \left(1- \frac{s}{m_{\tau}^2}\right)^2
  \left(1+ \frac{2s}{m_{\tau}^2}\right)
 Im \Pi_1(s+i\epsilon) \end{equation}
where the integration is over the invariant mass of the hadrons, s,
and we use the normalizations of ref. \cite{braaten2}.
$\Pi_1$ is the transverse part of the W boson current correlator
analogous to the Z boson current correlator of eq.(\ref{correlator})
(in eq.(\ref{correlator})
the neutral weak quark current should be replaced by the charged weak quark
current 
$ J^{\mu}=\frac{g}{2 \sqrt{2}}\sum_{i,j}
\overline{u_{i}}\gamma^{\mu} \left( 1-\gamma^{5} \right)
 V_{ij} d_{j} $ with $u_{i}= (u,c,t)$ and $d_{j}=(d,s,b)$).

We work in the approximation of massless $u,d,s$ quarks and a heavy
 $c$ quark within perturbative QCD omitting non-QCD corrections.
The $\alpha_s^3$ approximation in the leading order of a large mass expansion
was calculated in ref. \cite{gorishny2}. The first charm quark mass
suppressed term of the order $\alpha_s^2$ is obtained in ref. \cite{che}.
 We will obtain the large $c$ quark mass expansion of $R_{\tau}$
in the parameter
$\frac{m_{\tau}^2}{m_c^2}$ within effective QCD with 3 massless
flavours. This is the correct expansion
parameter since the mass of the $\tau$ lepton is
 below the threshold for the production of charmed hadrons.
Therefore a charm quark appears
only in internal fermion loops, the effective expansion parameter
appears to be $m_{\tau}^2/(4m_c^2)$ and a 
large c quark mass expansion is justified (although the charm quark mass
$m_c$ is smaller that the tau lepton mass $m_{\tau}$).
The Feynman diagrams that contribute to  $Im \Pi_1$ for the
W boson are of the non-singlet
type only (see figure 1) and were already calculated for the case of
Z boson decay (see eq.(\ref{mainvns})).
After performing the integration in eq.(\ref{rtau}) with the results of
 eq.(\ref{mainvns}) we have to apply the decoupling relation to go to 
effective QCD with 3 active flavours. Putting 
$\mu = m_{\tau}$ we obtain
\[
R_{\tau} = n \left( |V_{ud}|^2 + |V_{us}|^2 \right) 
\left[ 1+ \left( \frac{\alpha_{s}^{(3)}}{\pi} \right)  r_{1}^{0}  +
\left( \frac{\alpha_{s}^{(3)}}{\pi} \right)^{2}
\left( r_{2}^{0} + r_{2}^{1}\frac{m_{\tau}^2}{m_{c}^2}
  +r_{2}^{2}\frac{m_{\tau}^4}{m_{c}^4}
    +r_{2}^{3}\frac{m_{\tau}^6}{m_{c}^6} 
 + O( \frac{m_{\tau}^8}{m_{c}^8}) \right) \right. \hspace{5cm} \]
\begin{equation}
 \label{rtauresult}
 \left. \hspace{3cm} 
 + \left( \frac{\alpha_{s}^{(3)}}{\pi} \right)^{3}
\left( r_{3}^{0} + r_{3}^{1}\frac{m_{\tau}^2}{m_{c}^2}
  +r_{3}^{2}\frac{m_{\tau}^4}{m_{c}^4}
    +r_{3}^{3}\frac{m_{\tau}^6}{m_{c}^6}
   + O( \frac{m_{\tau}^8}{m_{c}^8}) \right)
   + O(\alpha_{s}^4)
\right] ,
 \end{equation}
$r^0_1 = C_F \left( \frac{3}{4} \right) $
\\ \\
$r^0_2 = 
       C_A CF  \left( \frac{947}{192} - \frac{11}{4} \zeta_3 \right)
      + C_F^2  \left(  - \frac{3}{32} \right)
      + n_f T_F C_F  \left(  - \frac{85}{48} + \zeta_3 \right) $
\\ \\
$r^1_2 = 
    T_F C_F \left[ \frac{107}{3000} - \frac{1}{150}
       \ln\left(\frac{m_{\tau}^2}{m_{c}^2}\right) \right] $ 
\\ \\
 $r^2_2 = 
     T_F C_F \left[  - \frac{1597}{5292000} 
       + \frac{1}{12600} \ln\left(\frac{m_{\tau}^2}{m_{c}^2}\right) \right] $
\\ \\
$r^3_2 = 
     T_F C_F \left[
 \frac{3991}{500094000} - \frac{1}{396900} 
      \ln\left(\frac{m_{\tau}^2}{m_{c}^2}\right) \right] $
\\ \\
$r^0_3 =
    n_f T_F C_A C_F \left(  - \frac{24359}{864} + \frac{73}{4}\zeta_3 
   + \frac{5}{6} \zeta_5 + \frac{11}{72} \pi^2 \right)
       + n_f T_F C_F^2 \left(  - \frac{125}{192} + \frac{19}{4} \zeta_3
                             - 5 \zeta_5 \right) $

$    + n_f^2 T_F^2 C_F \left( \frac{3935}{864} - \frac{19}{6} \zeta_3
           - \frac{1}{36} \pi^2 \right)
       + C_A C_F^2 \left(  - \frac{1733}{768} - \frac{143}{16} \zeta_3
            + \frac{55}{4} \zeta_5 \right) $

$       + C_A^2 C_F \left( \frac{559715}{13824} - \frac{2591}{96}\zeta_3
    - \frac{55}{24} \zeta_5 - \frac{121}{576} \pi^2 \right)
       + C_F^3 \left(  - \frac{69}{128} \right)  $
\\ \\
$r^1_3 =
        n_f T_F^2 C_F  \left[ - \frac{26333}{810000}  
                  +\frac{1}{1350} \pi^2
         - \frac{1}{250}\ln\left(\frac{m_{\tau}^2}{m_{c}^2}\right)
         + \frac{1}{450} \ln^2\left(\frac{m_{\tau}^2}{m_{c}^2}\right)
           \right] $

$       + T_F C_A C_F  \left[ - \frac{868427}{38880000}
     + \frac{53}{2880} \zeta_3  - \frac{11}{10800} \pi^2
         + \frac{31309}{648000} \ln\left(\frac{m_{\tau}^2}{m_{c}^2}\right)
         - \frac{11}{1200} \ln^2\left(\frac{m_{\tau}^2}{m_{c}^2}\right)
     \right] $

$       + T_F C_F^2  \left[  \frac{512251}{9720000}
               - \frac{529}{7200} \zeta_3 + \frac{11}{5400} \pi^2
          - \frac{581}{40500} \ln\left(\frac{m_{\tau}^2}{m_{c}^2}\right)
         + \frac{7}{1800} \ln^2\left(\frac{m_{\tau}^2}{m_{c}^2}\right)
     \right] 
       + T_F^2 C_F  \left( \frac{23}{3240}  \right) $
\\ \\
$r^2_3 =
      n_f T_F^2 C_F \left[ \frac{76567}{555660000}
          - \frac{1}{113400}\pi^2
         + \frac{37}{1984500}\ln\left(\frac{m_{\tau}^2}{m_{c}^2}\right)
   - \frac{1}{37800} \ln^2\left(\frac{m_{\tau}^2}{m_{c}^2}\right)
      \right] $

$       + T_F C_A C_F \left[  - \frac{93273701}{80015040000} +
         \frac{271}{138240} \zeta_3 - \frac{43}{1814400}\pi^2
      - \frac{424327}{381024000}\ln\left(\frac{m_{\tau}^2}{m_{c}^2}\right)
      + \frac{131}{604800} \ln^2\left(\frac{m_{\tau}^2}{m_{c}^2}\right)
      \right] $

$       + T_F C_F^2  \left[  - \frac{33317209}{6667920000}
              + \frac{2291}{2419200}\zeta_3
         + \frac{43}{907200} \pi^2
        + \frac{6887}{2976750} \ln\left(\frac{m_{\tau}^2}{m_{c}^2}\right)
        - \frac{23}{60480} \ln^2\left(\frac{m_{\tau}^2}{m_{c}^2}\right)
           \right] $

$       + T_F^2 C_F  \left[ \frac{767}{4860000}
             - \frac{1}{20250} \ln\left(\frac{m_{\tau}^2}{m_{c}^2}\right)
         \right] $
\\ \\
$r^3_3 =
     n_f T_F^2 C_F  \left[  - \frac{1094017}{472588830000} +
         \frac{1}{3572100} \pi^2
           - \frac{199}{750141000} \ln\left(\frac{m_{\tau}^2}{m_{c}^2}\right)
          + \frac{1}{1190700} \ln^2\left(\frac{m_{\tau}^2}{m_{c}^2}\right)
              \right] $

$       + T_F C_A C_F  \left[ \frac{13203894937}{16803158400000}
          - \frac{6217}{9676800} \zeta_3 - \frac{1}{142884000} \pi^2
    + \frac{260809}{17781120000}\ln\left(\frac{m_{\tau}^2}{m_{c}^2}\right)
    - \frac{73}{15876000} \ln^2\left(\frac{m_{\tau}^2}{m_{c}^2}\right)
             \right]  $

$     + T_F C_F^2  \left[ \frac{3881354513}{75614212800000}
             - \frac{121}{3763200}\zeta_3 + \frac{1}{71442000} \pi^2
   - \frac{10469}{240045120} \ln\left(\frac{m_{\tau}^2}{m_{c}^2}\right)
    + \frac{269}{23814000} \ln^2\left(\frac{m_{\tau}^2}{m_{c}^2}\right)
                 \right]  $

$       + T_F^2 C_F  \left[  - \frac{16661}{2500470000}
            + \frac{1}{441000} \ln\left(\frac{m_{\tau}^2}{m_{c}^2}\right)
          \right] $, \\ \\
where $\alpha_s^{(3)}\equiv \alpha_s^{(3)}(m_{\tau})$, 
 $m_c\equiv m_c(m_{\tau})$ is the $\overline{MS}$ charm quark mass
and n = 3 is the number of quark colours.
We neglected terms that are suppressed  by the bottom and top quark masses
(which in principle are present after the decoupling).
Substitution of the QCD colour factors and $n_f$=3 gives \\ \\
$r^0_1 = 1,$\hspace{1cm} 
$r^0_2 = 5.2023 ,$\\
$r^1_2 = 0.023778 - 0.0044444 \ln\left(\frac{m_{\tau}^2}{m_{c}^2}\right) ,$\\
$r^2_2 =  - 0.00020118 + 0.000052910 \ln\left(\frac{m_{\tau}^2}{m_{c}^2}\right)
   ,$ \\
$r^3_2 = 0.0000053203 
    - 0.0000016797 \ln\left(\frac{m_{\tau}^2}{m_{c}^2}\right), $\\
$r^0_3 = 26.3659, $\\ 
$r^1_3 = - 0.057156 + 0.079881 \ln\left(\frac{m_{\tau}^2}{m_{c}^2}\right)
      - 0.012654 \ln^2\left(\frac{m_{\tau}^2}{m_{c}^2}\right),$ \\
$r^2_3 = - 0.00099668  - 0.00016858 \ln\left(\frac{m_{\tau}^2}{m_{c}^2}\right)
   + 0.0000687096 \ln^2\left(\frac{m_{\tau}^2}{m_{c}^2}\right),$ \\ 
 $r^3_3 = 0.000036522 - 0.0000089407
  \ln\left(\frac{m_{\tau}^2}{m_{c}^2}\right)
  + 0.00000168435 \ln^2\left(\frac{m_{\tau}^2}{m_{c}^2}\right),$ 
\\ \\
Note that the coefficient $r^1_2$ agrees with ref. \cite{che}.

Although the expansion parameter
 $\frac{m_{\tau}^2}{m_{c}^2} \approx \left(\frac{1.777}{1.3\pm 0.3}\right)^2$
is slightly larger than 1, the fast decrease of the coefficients ensures
a good convergence of the large charm quark mass expansion.

We conclude that the large mass expansion converges fast for both
Z boson and $\tau$ lepton decays and the obtained $\alpha_s^3$ approximations
can be trusted.

\section{Acknowledgements}
We are grateful to D.Yu. Bardin,
 D.J. Broadhurst, K.G. Chetyrkin, A.I. Davydychev,
 A.L. Kataev, R. Kleiss and P.J. Nogueira for helpful discussions.
 One of us, S.L., is grateful to the \nopagebreak
theory group of HIKHEF-H for its kind hospitality.
This work is supported in part by INTAS, Grant no. 93-1180.
\\ \\

\end{document}